\documentclass[slac_one]{revtex4}
\usepackage{graphicx}
\usepackage{fancyhdr}
\def\Journal#1#2#3#4{{#1} {\bf #2}, #3 (#4)}

\def\beq{\begin{equation}}
\def\eeq{\end{equation}}
\def\bea{\begin{array}}
\def\eea{\end{array}}
\def\be{\begin{equation}}
\def\ee{\end{equation}}
\def\ba{\begin{eqnarray}}
\def\ea{\end{eqnarray}}

\def\to{\rightarrow}

\def\[{\left[}
\def\]{\right]}
\def\({\left(}
\def\){\right)}

\def\sm0{{\widetilde{m}_0}}

\def\U1em{{U(1)_{\rm em}}}
\def\to{\rightarrow}

\def\sq2{\sqrt{2}}

\def\ee{e^+e^-}

\def\End{\end{document}}

\def\NPB{{\em Nucl. Phys.} B}
\def\PLB{{\em Phys. Lett.}  B}
\def\PRL{\em Phys. Rev. Lett.}
 
\def\PRD{{\em Phys. Rev.} D}

\begin{document}

\title{{\small{2005 International Linear Collider Workshop - Stanford,
U.S.A.}}\\ 
\vspace{14pt}
Measurement of lepton flavor violating Yukawa couplings at ILC}

\author{S. Kanemura\footnote{Speaker}, T.~Ota, and K.~Tsumura}

\affiliation{
Department of Physics, Osaka University, Toyonaka, 
Osaka 560-0043, Japan}

\begin{abstract}

We discuss 
lepton flavor violation (LFV) associated with tau leptons 
in the general framework of the two Higgs doublet 
model, in which LFV couplings are introduced as a deviation 
from Model II Yukawa interaction.
Parameters of the model are constrained from experimental results 
and also from requirements of theoretical consistencies 
such as vacuum stability and perturbative unitarity.
Current data for LFV rare tau decays provide substantial 
upper limits on the LFV Yukawa couplings 
in the large $\tan\beta$ region,  
which are comparable with predictions in fundamental 
theories. Here $\tan\beta$ is the ratio of vacuum 
expectation values of the two Higgs doublets.
A search for the LFV decays 
$\phi^{0} \rightarrow \tau^\pm \mu^\mp$ $(\tau^\pm e^\mp)$ 
of neutral Higgs bosons ($\phi^{0} =h,H$ and $A$) 
at future collider experiments
can be useful to further constrain the LFV couplings, 
especially in the relatively small $\tan\beta$ region 
($\tan\beta \lesssim 30 $),  
where rare tau decay data cannot give any strong limit.

\end{abstract}

\maketitle

\thispagestyle{fancy}


\section{Introduction}

The structure of the electroweak symmetry beaking sector 
would directly connect to the property of physics beyond 
the standard model (SM).
Models of such new physics predict extended Higgs sectors
with more than one scalar doublets in the low energy effective theory. 
These extended Higgs models would show distinctive features from the SM
phenomenology. 
The most obvious evidence would be 
the confirmation of the existence of the extra scalar
states such as CP-odd and charged states at future collider 
experiments. 
Even if they are too heavy to be directly discovered   
and only the lightest Higgs boson is found at the experiments, 
we would be able to explore the extended Higgs sector by searching for 
deviations from the SM predictions on the couplings with gauge 
bosons and fermions as well as on the self coupling. 
Furthermore, search for non-SM interactions 
can also be useful.

Lepton flavor violation (LFV) is an example 
for such non-SM phenomena. 
In particular, LFV in the Yukawa sector can only appear 
for extended Higgs sectors. 
Flavor violation between electrons and muons\cite{KunoOkada}   
has been tested through rare muon decays such as  
$\mu^{}\rightarrow e^{} \gamma$ and  
$\mu^{} \rightarrow e^{}e^{+}e^{-}$,  
as well as through $\mu$-$e$ conversion.  
Tau lepton associated LFV has also been studied by rare decays of 
tau leptons such as 
$\tau \rightarrow \ell_i P^0 $\cite{belle-tau-lp0},
$\tau \rightarrow \ell_i M^+M'^-$\cite{belle-tau-lMM,babar-tau-lMM}, 
$\tau \rightarrow \ell_i\ell'^+\ell'^-$\cite{belle-tau-3l,babar-tau-3l}, and   
$\tau \rightarrow \ell_i\gamma$\cite{belle-tau-egamma,
belle-tau-mugamma,babar-tau-mugamma},
where $\ell_i$ ($i=1,2$) respectively represent an electron and a muon, 
$P^0$ does $\pi^0$, $\eta$ and $\eta'$ mesons, 
$M^\pm$ ($M'^\pm$) does $\pi^\pm$ and $K^\pm$ mesons, 
and $\ell'^\pm=e^\pm$ and $\mu^\pm$.
The LFV Yukawa couplings can be constrained from 
the data for these processes especially those with the Higgs boson mediation. 
For $\mu$-$e$ mixing, the Higgs boson mediated LFV coupling 
has been discussed in Ref.~\cite{KitanoKoikeKomineOkada,muegamma-type3THDM}. 
Tau lepton associated LFV processes with the Higgs boson mediation 
have been discussed in models with 
supersymmetry (SUSY)\cite{BK,DER,Sher-tmeta,Rossi}
as well as in the two Higgs doublet model (THDM) 
in some specific scenarios\cite{cheng-sher,HiggsLFV-THDM,Iltan}.
In Ref.~\cite{BHHS}, tau associated LFV processes have 
been discussed comprehensively 
in the framework of 4-Fermi contact interactions. 
Phenomenological consequences of the LFV Yukawa couplings 
associated with tau leptons have also been studied 
for future observables 
such as $B_s$ decays\cite{bsmutau,DER} at 
(super) B factories\cite{super-B} and 
Higgs boson decays\cite{Pilaftsis,Rossi,Herrero,kot} 
at CERN LHC\cite{Assamagan}, 
an electron-positron linear collider (LC)\cite{Osaka} 
and a muon collider\cite{muon-collider}.
Furthermore, deep inelastic scattering processes 
$\mu N \to \tau X$\cite{Sher-turan,mutauDIS} from 
intense high energy muons at neutrino factories (or muon colliders)
and $e N \to \tau X$\cite{mutauDIS} by using the electron (positron) 
beam of a LC would be useful to further explore 
the tau lepton associated LFV Yukawa couplings. 

In this talk, we discuss LFV in Higgs boson decays  
into a $\tau$-$\ell_i$ pair 
in the general framework of the THDM\cite{kot}. 
We study experimental upper limits 
on the tau lepton associated LFV Yukawa couplings  
to evaluate possible maximal values of the branching fractions.
The parameter space of the model is 
tested by theoretical requirements for vacuum stability\cite{VS} 
and perturbative unitarity\cite{PU-1,PU-2}.
Current data from electroweak precision measurements 
at LEP\cite{LEP,smlike1} 
and those at the B factories\cite{bsg-ex} 
also strongly constrain parameters of the Higgs potential. 
Under these theoretical bounds and experimental limits on the model, 
possible maximal values of the
LFV couplings of $\tau$-$\ell_i$-$\phi^{0}$ are obtained 
by using the current data for rare tau decays, where 
$\phi^{0}$ represents two CP-even ($h$ and $H$) 
and a CP-odd ($A$) Higgs bosons. 
We then evaluate branching ratios of $\phi^0 \to \tau^\pm \ell_i^\mp$ 
with the maximal allowed values of the 
LFV couplings of $\tau$-$\ell_i$-$\phi^{0}$ 
in a wide range of the parameter space.

\section{Lepton flavor violation in Yukawa interaction}
\label{Sec:Model}

The Higgs sector of the general THDM is expressed as
\begin{eqnarray}
-\mathcal{L}_{\text{Higgs}} 
&=&
m_{1}^{2} \left| \Phi_{1} \right|^{2}
+
m_{2}^{2} \left| \Phi_{2} \right|^{2}
-
\left( m_{3}^{2} \Phi_{1}^{\dagger} \Phi_{2} + {\rm H.c.} \right)
+
\frac{\lambda_{1}}{2} \left|\Phi_{1}\right|^{4}
+
\frac{\lambda_{2}}{2} \left|\Phi_{2}\right|^{4} 
+
\lambda_{3} \left|\Phi_{1}\right|^{2} \left|\Phi_{2}\right|^{2}
+
\lambda_{4} \left|\Phi_{1}^{\dagger} \Phi_{2} \right|^{2}\nonumber \\
&&
+
\left\{
 \frac{\lambda_{5}}{2}
 \left( \Phi_{1}^{\dagger} \Phi_{2} \right)^{2} + {\rm H.c.}
\right\} + \left(
 \lambda_{6} \left|\Phi_{1}\right|^{2}
 \Phi_{1}^{\dagger}
 \Phi_{2} +{\rm H.c.} \right) 
 + \left(
 \lambda_{7} \left|\Phi_{2}\right|^{2}
 \Phi_{1}^{\dagger}
 \Phi_{2} +{\rm H.c.} \right), 
\label{eq:LHiggs}
\end{eqnarray}
where $\Phi_{1}$ and $\Phi_{2}$ are 
the scalar iso-doublets with hypercharge $1/2$.
In Eq.~(\ref{eq:LHiggs}), 
$m_{3}^{2}$, $\lambda_{5}$, $\lambda_{6}$ and
$\lambda_{7}$ are complex in general.
We here assume that all the parameters $m_{1-3}^{2}$ and $\lambda_{1-7}$ are
real. 
The terms of $m_{3}^{2}$, $\lambda_{6}$ and
$\lambda_{7}$ break the discrete symmetry explicitly. 
As we consider the model in which the discrete symmetry is explicitly 
broken only in the leptonic Yukawa interaction, 
we set the hard-breaking coupling constants 
to be zero in the Higgs potential; 
i.e., $\lambda_{6} = \lambda_{7} =0$\footnote{%
Even in such a case, $\lambda_{6}$ and $\lambda_{7}$
are effectively induced at the loop level.
They are suppressed by the loop factor, so that we here neglect
these small effects.}, 
and retain only the soft-breaking mass parameter $m_3^2$.

There are eight degrees of freedom in the two Higgs doublet fields.
Three of them are absorbed by the weak gauge bosons via the Higgs
mechanism, and remaining five are physical states.
After the diagonalization of the mass matrices, they correspond to 
two CP-even ($h$ and $H$), 
a CP-odd ($A$), and a pair of charged  ($H^{\pm}$) Higgs bosons.
We define such that $h$ is lighter than $H$.
The eight real parameters $m_{1-3}^{2}$ and $\lambda_{1-5}$ 
can be described by the same number of physical parameters; i.e.,
the vacuum expectation value $v$ $(\simeq 246$ GeV), 
the Higgs boson masses 
$m_{h}^{}, m_{H}^{}, m_{A}^{}$ and  $m_{H^{\pm}}^{}$,  
the mixing angle $\alpha$ between the CP-even Higgs bosons, 
the ratio $\tan \beta$ 
($ \equiv \langle \Phi^{0}_{2} \rangle / \langle \Phi^{0}_{1}
\rangle$) 
of the vacuum expectation values for two Higgs doublets,
and the soft-breaking scale $M$ ($\equiv \sqrt{m_{3}^{2}/\sin\beta
\cos\beta}$) for the discrete symmetry.
The quartic couplings are expressed in terms of physical parameters 
in Ref.~\cite{smlike2} 

Parameters of the Higgs sector are constrained from 
requirements of theoretical consistencies 
and also from the current experimental results. 
We here take into account two kinds of theoretical conditions; i.e.,  
vacuum stability\cite{VS} and perturbative 
unitarity\cite{PU-1,PU-2} at the tree level.
The condition for tree-level unitarity,
which we employ in Refs.~\cite{PU-2,smlike2}, 
is described as
$\left|\langle \phi_{3} \phi_{4}  
| a^{0}  | \phi_{1} \phi_{2} \rangle   \right|
< \xi, $
where
$\langle \phi_{3} \phi_{4}  |a^{0}  | \phi_{1} \phi_{2} \rangle $ is the 
$s$-wave
amplitude for the process of $\phi_{1} \phi_{2} \rightarrow \phi_{3}
\phi_{4}$ with 
$\phi_{a}$ ($a$=1-4) denoting Higgs bosons 
and longitudinal components of weak gauge bosons.
We take the criterion $\xi$ to be 1
(and also 1/2 for comparison).
The experimental constraints are provided by the
LEP precision data\cite{LEP}, 
the $b \rightarrow s \gamma$ results\cite{bsg-ex}, 
and the direct search results for the Higgs bosons\cite{LEP}.  
The LEP precision data provide the strong constraint 
on the new physics structure via the gauge-boson two-point functions. 
The constraint on $\rho$ parameter indicates that the Higgs sector 
is approximately custodial $SU(2)$ symmetric.
This requirement is satisfied when 
(i) $m_{H^{\pm}}^{} \simeq m_{A}^{}$,  
(ii) $m_{H^{\pm}}^{} \simeq m_{H}^{}$ with
$\sin^{2} (\alpha - \beta)\simeq 1$, and 
(iii) $m_{H^{\pm}}^{} \simeq m_{h}^{}$ with
$\cos^{2} (\alpha - \beta) \simeq 1$.
It is known that in Model II,  the $b \rightarrow s \gamma $ result 
gives the lower bound on the charged Higgs boson mass.
We here take into account this bound by requiring
$m_{H^{\pm}}^{} \gtrsim 350$ GeV.

Next, 
we consider the Yukawa interaction for charged leptons as   
\begin{eqnarray}
-\mathcal{L}_{\text{lepton}}=
\overline{\ell}_{R i} 
 \left\{
 Y_{\ell_{i}} \delta_{ij} \Phi_{1}  
 +
 \left(
 Y_{\ell_{i}} \epsilon^{L}_{ij} 
 +
 \epsilon^{R}_{ij} Y_{\ell_{j}}
 \right) \Phi_{2} 
 \right\}
 \cdot
 L_{j} + {\rm H.c.},
\label{eq:Llepton}
\end{eqnarray}
where 
$\ell_{R i}$ ($i$=1-3) are 
right-handed charged leptons, and 
$L_{i}$ ($i$=1-3) denote the lepton doublets 
and $Y_{\ell_{i}} $ are the Yukawa
couplings for $\ell_{i}$. 
This interaction is reduced to be of Model II\cite{HHG}  
in the limit $\epsilon^{L,R}_{ij} \rightarrow 0$ with 
the discrete symmetry under $e_{R}^{i} \rightarrow + e_{R}^{i}$,
$L_{i} \rightarrow + L_i$,
$\Phi_{1} \rightarrow + \Phi_{1}$, and
$\Phi_{2} \rightarrow - \Phi_{2}$.   
Nonzero values of $\epsilon_{ij}^{L,R}$ ($i\neq j$)
yield the LFV Yukawa couplings after the diagonalization 
of the mass matrix.
We note that in supersymmetric standard models, 
the Yukawa interaction for leptons is of Model II 
at the tree level, and $\epsilon_{ij}^{L,R}$ 
can be induced at the loop level due to slepton 
mixing\cite{MSSMRN,BK,DER}.
For the quark sector, Model II Yukawa interactions are assumed to 
suppress FCNC, 
imposing the invariance under the transformation of 
$u_{R}^{i} \rightarrow - u_{R}^{i}$, $d_{R}^{i} \rightarrow + d_{R}^{i}$,
$q_{L}^{i} \rightarrow + q_{L}^{i}$,
$\Phi_{1} \rightarrow + \Phi_{1}$, and
$\Phi_{2} \rightarrow -\Phi_{2}$.

The tau lepton associated LFV interactions 
in Eq.~(\ref{eq:Llepton}) can be reduced 
in the mass eigenbasis of each field to
\begin{eqnarray}
-\mathcal{L}_{\tau{\rm LFV}}
&=&
\frac{m_{\tau}}{v \cos^{2}\beta}
 \left(
 \kappa^{L}_{3i} \overline{\tau} {\rm P}_{L} \ell_{i}
 +
 \kappa^{R}_{i3} \overline{\ell}_i {\rm P}_{L} \tau
 \right)
 \left\{
 \cos\left(\alpha - \beta \right) h
 +
 \sin\left(\alpha - \beta \right) H
 -
 {\rm i} A
 \right\} \nonumber \\
& &+
 \frac{\sqrt{2} m_{\tau}}{v \cos^{2}\beta}
 \left(
 \kappa^{L}_{3i}
 \overline{\tau} {\rm P}_{L} \nu_{i} 
 +
 \kappa^{R}_{i3}
 \overline{\ell_{i}} {\rm P}_{L} \nu_{\tau}
 \right)  H^{-}
 +
 {\rm H.c.},
\label{eq:tauLFV}
\end{eqnarray}
where ${\rm P}_{L}$ is the projection operator to 
the left-handed field, and 
$\ell_1$ and $\ell_2$ respectively represent $e$ and $\mu$.
In general, the LFV parameters $\kappa^{L,R}_{ij}$ can be expressed in
terms of $\epsilon^{L,R}_{ij}$ and $\tan\beta$.\footnote{%
LFV parameters $\kappa_{ij}^{L,R}$ can be expressed as 
\begin{eqnarray}
\kappa^{X}_{ij} 
= 
-\frac{\epsilon^{X}_{ij}}{
 \left\{ 1+ (\epsilon^{L}_{33} + \epsilon^{R}_{33})\tan\beta
 \right\}^{2} }
\qquad (X=L,R),
\end{eqnarray}
for the case of 
$\epsilon^{L,R}_{ij} \tan\beta \ll \mathcal{O}(1)$ 
which is satisfied in the case of the 
minimal supersymmetric standard model (MSSM)\cite{BK,DER}.}
We here take these $\kappa^{L,R}_{ij}$ as effective couplings,
and investigate their phenomenological consequences.
We note that Eq.~(\ref{eq:tauLFV}) is exact in the limit of 
$m_{\ell_{i}}^{} \to 0$.
The terms of $\kappa_{i3}^L$ and $\kappa_{3i}^R$ ($i=1, 2$) 
are proportional to $m_{\ell_i}^{}$, 
so that they decouple in this approximation.

We briefly discuss relationship between 
$\kappa^{L,R}_{ij}$ and new physics models beyond the
cut-off scale of the {\it effective} THDM.
When a new physics model is specified at the high energy scale, 
$\kappa^{L,R}_{ij}$ can be predicted as a function of the model 
parameters.
For example, in the MSSM, slepton mixing can be a source of LFV.
Notice that the induced LFV Higgs interactions do not necessarily 
decouple in the limit where the SUSY  particles are sufficiently heavy, 
because their couplings only depend on the ratio of the SUSY 
parameters.
Therefore, the Higgs associated LFV processes can become
important in a scenario with the soft-SUSY-breaking scale 
$m_{\rm SUSY}^{}$ to be much higher than the electroweak one. 
In the MSSM, predicted values of $|\kappa_{3i}^L|^2$ 
can be as large as of ${\cal O}({10^{-6}})$ when 
$m_{\rm SUSY}^{}$ is a few TeV\cite{Rossi,Osaka}.
In the MSSM with right-handed neutrinos, 
left-handed slepton mixing may be a consequence of running effects of the
neutrino Yukawa couplings between the scale of the grand unification 
and that of the right-handed neutrinos\cite{MSSMRN}.
The parameters $\kappa^{L}_{3i}$ are mainly induced by mixing 
of left-handed sleptons\cite{BK,DER,Sher-tmeta,Rossi,Osaka,Herrero}.
The LFV Yukawa interactions can also appear effectively in the Zee 
model\cite{Zee}.
The LFV parameters $\kappa^{L,R}_{ij}$ are induced through flavor
violating couplings in the charged scalar interactions with leptons.

\begin{table}[t] {
\begin{tabular}{lll} 
\hline \hline
Mode &    Belle (90\% CL) \hspace*{1cm}& BaBar (90\% CL)  \\
\hline 
$\tau^-\rightarrow e^-\pi^0$&
\underline{$1.9\times 10^{-7}$}\cite{belle-tau-lp0}&  \\
$\tau^-\rightarrow e^-\eta$&
\underline{$2.4\times 10^{-7}$}\cite{belle-tau-lp0}&  \\
$\tau^-\rightarrow e^-\eta'$&
\underline{$10\times 10^{-7}$}\cite{belle-tau-lp0}&  \\
$\tau^-\rightarrow \mu^-\pi^0$&
\underline{$4.1\times 10^{-7}$}\cite{belle-tau-lp0}&  \\
$\tau^-\rightarrow \mu^-\eta$&
\underline{$1.5\times 10^{-7}$}\cite{belle-tau-lp0}&  \\
$\tau^-\rightarrow \mu^-\eta'$&
\underline{$4.7\times 10^{-7}$}\cite{belle-tau-lp0}&  \\
\hline
$\tau^-\rightarrow e^-\pi^+\pi^-$&
$8.4\times 10^{-7}$\cite{belle-tau-lMM}&
\underline{$1.2\times 10^{-7}$}\cite{babar-tau-lMM}  \\
$\tau^-\rightarrow e^-\pi^+K^-$&
$5.7\times 10^{-7}$\cite{belle-tau-lMM}&
\underline{$3.2\times 10^{-7}$}\cite{babar-tau-lMM}  \\
$\tau^-\rightarrow e^-K^+\pi^-$&
$5.6\times 10^{-7}$\cite{belle-tau-lMM}&
\underline{$1.7\times 10^{-7}$}\cite{babar-tau-lMM}  \\
$\tau^-\rightarrow e^-K^+K^-$&
$3.0\times 10^{-7}$\cite{belle-tau-lMM}&
\underline{$1.4\times 10^{-7}$}\cite{babar-tau-lMM}  \\
$\tau^-\rightarrow \mu^-\pi^+\pi^-$&
\underline{$2.8\times 10^{-7}$}\cite{belle-tau-lMM}&
$2.9\times 10^{-7}$\cite{babar-tau-lMM}  \\
$\tau^-\rightarrow \mu^-\pi^+K^-$&
$6.3\times 10^{-7}$\cite{belle-tau-lMM}&
\underline{$2.6\times 10^{-7}$}\cite{babar-tau-lMM}  \\
$\tau^-\rightarrow \mu^-K^+\pi^-$&
$15.5\times 10^{-7}$\cite{belle-tau-lMM}&
\underline{$3.2\times 10^{-7}$}\cite{babar-tau-lMM}  \\
$\tau^-\rightarrow \mu^-K^+K^-$&
$11.7\times 10^{-7}$\cite{belle-tau-lMM}&
\underline{$2.5\times 10^{-7}$}\cite{babar-tau-lMM}  \\
\hline
$\tau^-\rightarrow e^-e^+e^-$\hspace*{1cm}&
$3.5\times10^{-7}$\cite{belle-tau-3l}&
\underline{$2.0\times10^{-7}$}\cite{babar-tau-3l}  \\
$\tau^-\rightarrow e^-\mu^+\mu^-$&
\underline{$2.0\times 10^{-7}$}\cite{belle-tau-3l}&
$3.3\times 10^{-7}$\cite{babar-tau-3l}  \\
$\tau^-\rightarrow \mu^-e^+e^-$&
\underline{$1.9\times 10^{-7}$}\cite{belle-tau-3l}&
$2.7\times 10^{-7}$\cite{babar-tau-3l}  \\
$\tau^-\rightarrow \mu^-\mu^+\mu^-$&
$2.0\times 10^{-7}$\cite{belle-tau-3l}&
\underline{$1.9\times 10^{-7}$}\cite{babar-tau-3l}  \\
\hline
$\tau\rightarrow e\gamma$&
\underline{$3.9\times 10^{-7}$}\cite{belle-tau-egamma}&  \\
$\tau\rightarrow \mu\gamma$&
$3.1\times 10^{-7}$\cite{belle-tau-mugamma}& 
\underline{$6.8\times 10^{-8}$}\cite{babar-tau-mugamma}   \\
\hline \hline
\end{tabular}}
\caption{Current experimental limits 
on branching ratios of the LFV rare tau decays.}
\label{Tab:tau-bound}
\end{table}

\section{Bound on LFV Yukawa couplings from
 rare tau decays} 
\label{Sec:LFV-tau-decay}

In order to constrain the LFV parameters 
$|\kappa_{3i}^{L}|$ and $|\kappa_{i3}^{R}|$,
we take into account the data for rare tau decay processes 
such as 
$\tau \rightarrow \ell_i P^0 $,
$\tau \rightarrow \ell_i M^+M'^-$, 
$\tau \rightarrow \ell_i\ell'^+\ell'^-$, and   
$\tau \rightarrow \ell_i\gamma$,
where $P^0$ represents $\pi^0$, $\eta$ and $\eta'$ mesons, 
$M^\pm$ ($M'^\pm$) does $\pi^\pm$ and $K^\pm$ mesons, 
and $\ell'^\pm=e^\pm$ and $\mu^\pm$.
The list of the current data from the B factories are 
summarized in Table~\ref{Tab:tau-bound}
\cite{belle-tau-lp0,belle-tau-lMM,babar-tau-lMM,
belle-tau-3l,babar-tau-3l,belle-tau-egamma,
belle-tau-mugamma,babar-tau-mugamma}. 
These bounds may be improved at the super B factory
around a digit\cite{super-B}. 
In our analysis, we take the underlined data in
Table~\ref{Tab:tau-bound} as our numerical inputs. 
The branching ratios for these rare $\tau$ LFV decays are 
summarized in Ref.~\cite{kot}.

Since branching ratios for $\tau^-\to\ell_i^-P^0$,
$\tau^-\to\ell_i^-M^+M^-$, $\tau^-\to\ell_i^-\ell'^+\ell'^-$ and 
$\tau^-\to\ell_i^-\gamma$ 
depend on different combinations of the Higgs boson masses,
independent information can be obtained for the model parameters 
by measuring each of them. When all the masses of Higgs bosons are large,
these decay processes decouple by a factor of $1/m_{\rm Higgs}^{4}$.
Although these branching ratios are complicated functions of the mixing angles,
each of them can be simply expressed to be proportional to $\tan^{6} \beta$ for
$\tan\beta \gg 1$ in the
SM-like region ($\sin(\alpha-\beta) \sim -1$ \cite{GH,smlike1,smlike2}).
This $\tan^{6} \beta$ dependence is a common feature of the tau-associated
LFV processes with the Higgs-mediated 4-Fermi interactions.

The experimental upper limit 
on $|\kappa_{3i}|^{2} (\equiv |\kappa_{3i}^L|^{2}+|\kappa_{i3}^R|^{2})$ 
can be obtained  
by using the experimental results given 
in Table~\ref{Tab:tau-bound} and analytic
expressions of the decay branching ratios 
for rare tau LFV decay processes in Ref.~\cite{kot}.
For description, let us consider the bound 
from the $\tau\rightarrow \mu\eta$ results\cite{Sher-tmeta}; 
\begin{eqnarray}
&|\kappa_{32}|^{2} \leq
 \left( 
 \left|
  \kappa^{\text{max}}_{32}
 \right|^{2}
 \right)_{\tau^{}\rightarrow \mu^{}\eta}
 \equiv
 \frac{256 \pi {\rm Br}(\tau^{} \rightarrow \mu^{} \eta)_{\rm exp}  m_{A}^{4}}
 {9 G_{F}^{2} m_{\tau}^{3} m_{\eta}^{4} F_{\eta}^{2} \tau_{\tau} 
  \left(
   1- \frac{m_{\eta}^{2}}{m_{\tau}^{2}}
  \right)^{2}}
 \frac{\cos^{6}\beta}{\sin^{2}\beta},
\label{eq:bound-from-tmeta}
\end{eqnarray}
where ${\rm Br}(\tau^{} \rightarrow \mu^{} \eta)_{\rm exp}$ is the
experimental upper limit on the branching ratio of $\tau^{} \rightarrow
\mu^{} \eta$ in Table~\ref{Tab:tau-bound}.
In particular, for $\tan\beta\gg1$,
the right-hand-side can be expressed by
\begin{eqnarray}
 \left( 
 \left|
  \kappa^{\text{max}}_{32}
 \right|^{2}
 \right)_{\tau^{}\rightarrow \mu^{}\eta}
\simeq
2.3 \times 10^{-4} \times 
\left(
 \frac{m_{A}^{}}{350 {\rm [GeV]}}
\right)^{4}
\left(
\frac{30}{\tan\beta}
\right)^{6}.
\label{eq:kappa32sqmax}
\end{eqnarray} 
It can be easily seen that the bound 
$(|\kappa_{32}^{\text{max}}|^{2})_{\tau\rightarrow \mu\eta}$ 
is rapidly relaxed 
in the region with small $\tan\beta$ and large $m_{A}^{}$. 
In a similar way to  Eq.~(\ref{eq:bound-from-tmeta}),
the maximal allowed value
$\left( |\kappa_{3i}^{\text{max}}|^{2} \right)_{\text{mode}}$ can be
calculated for each mode.
The combined upper limit $|\kappa^{\text{max}}_{3i}|^{2}$ is then given by
\begin{eqnarray}
\left|\kappa^{\text{max}}_{3i}\right|^{2}
 \equiv
 {\rm MIN}
 \left\{
  \left(
  \left|
   \kappa^{\text{max}}_{3i}
  \right|^{2}
  \right)_{\tau\rightarrow\ell\eta},
  \left(
  \left|
   \kappa^{\text{max}}_{3i}
  \right|^{2}
  \right)_{\tau \rightarrow \ell \mu^{+}\mu^{-}},
  \left(
  \left|
   \kappa^{\text{max}}_{3i}
  \right|^{2}
  \right)_{\tau \rightarrow \ell K^{+}K^{-}},
  \left(
  \left|
   \kappa^{\text{max}}_{3i}
  \right|^{2}
  \right)_{\tau \rightarrow \ell \gamma}, \cdot\cdot\cdot
 \right\}.
\label{eq:kappaSq-max}
\end{eqnarray}
As shown below, 
$\tau^{}\rightarrow \ell^{}_i\eta$ and  
$\tau^{}\rightarrow \ell^{}_i\gamma$
give the strongest upper limits on $|\kappa_{3i}|^{2}$
in a wide range of the parameter space. 
In addition, in some parameter regions  
$\tau^{}\rightarrow \ell^{}_iK^{+}K^{-}$
and $\tau^{}\rightarrow \ell^{}_i\mu^{+}\mu^{-}$ can 
also give similar limits on $|\kappa_{3i}|^{2}$ 
to those from the above two processes.

\begin{figure}[t]
\begin{minipage}{8cm}
\unitlength=1cm
\begin{picture}(7.8,6)
\put(0,1){\includegraphics[width=8cm]{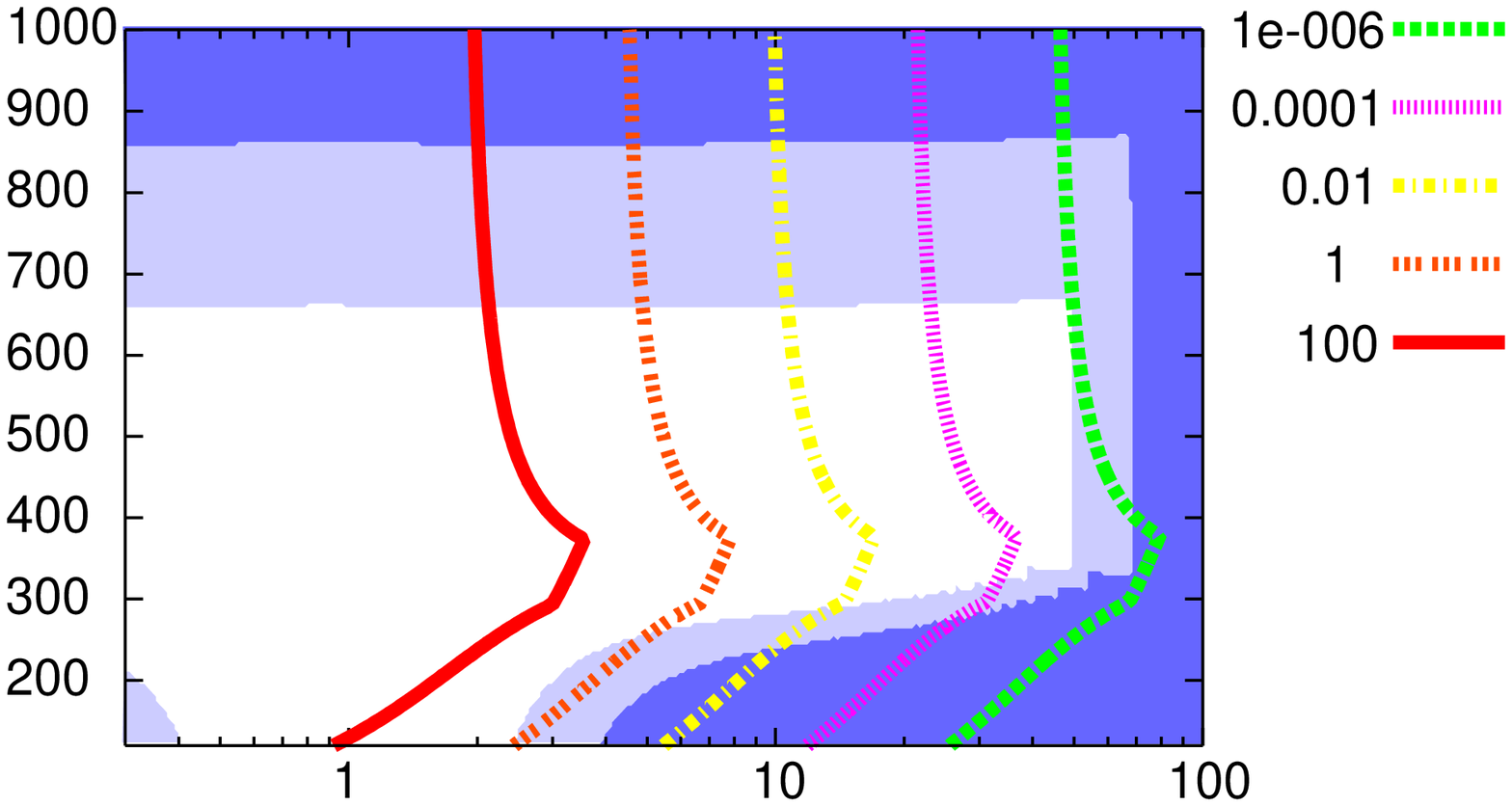}}
\put(0,5.5){$m_{A}^{}$ [GeV]}
\put(6.8,1){$\tan\beta$}
\put(3,0.5){(a)}
\end{picture}
\end{minipage}
\begin{minipage}{8cm}
\unitlength=1cm
\begin{picture}(7.8,6)
\put(0,1){\includegraphics[width=7.8cm]{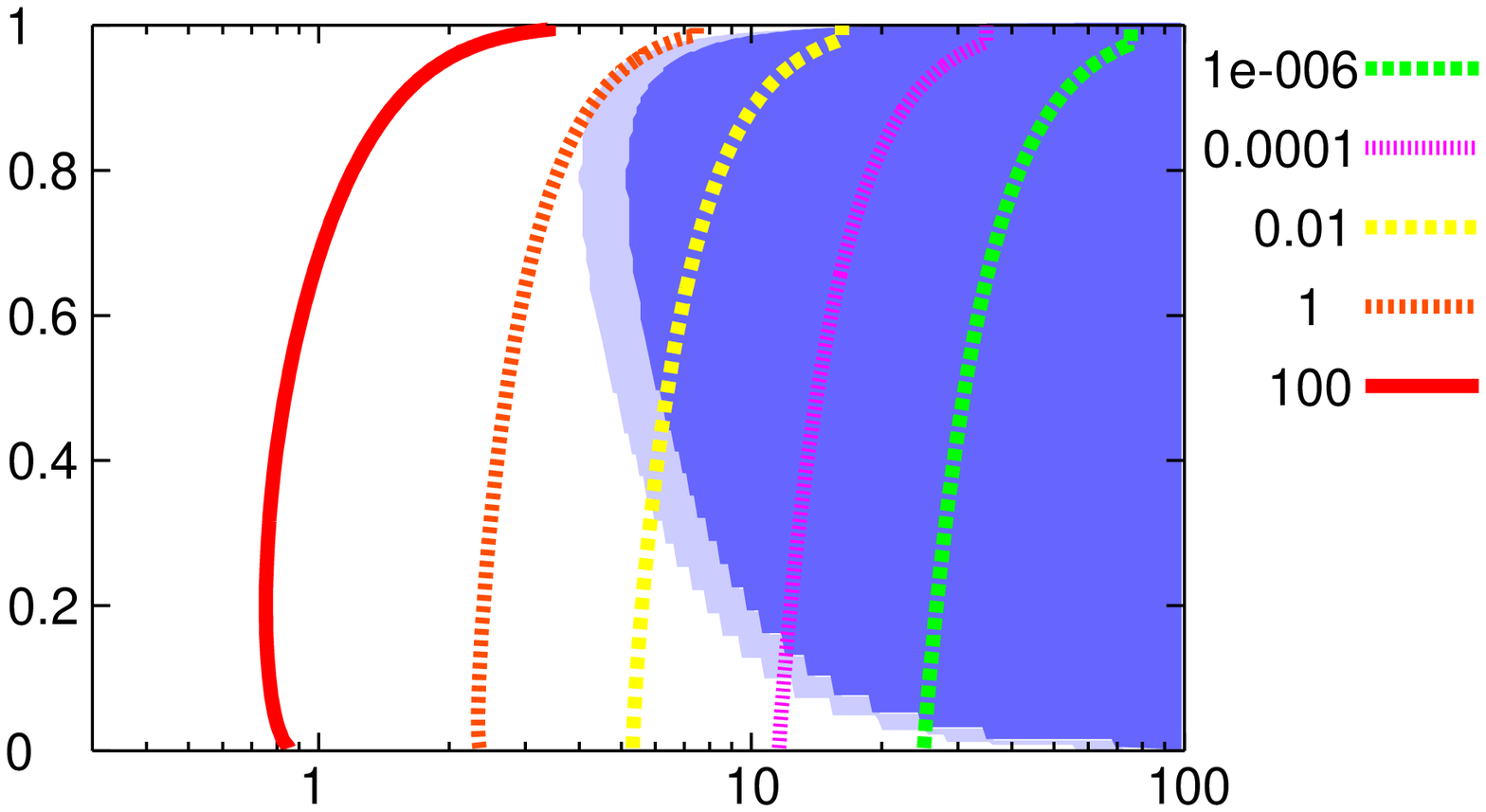}}
\put(0,5.6){$\sin^{2}(\alpha-\beta)$}
\put(7,1){$\tan\beta$}
\put(3,0.5){(b)}
\end{picture}
\end{minipage}
\caption{
Contours of  $|\kappa^{\text{max}}_{32}|^{2}$,  
the possible maximal value of $|\kappa_{32}|^{2}$ from the rare tau
decay results, are shown (a) in the $\tan\beta$-$m_{A}$ plane and (b) 
in the $\tan\beta$-$\sin^{2}(\alpha-\beta)$ plane.
The parameters are taken to be (a) $m_{h}^{} =120$ GeV, 
$m_{H}^{} = m_{H^{\pm}} = 350$ GeV and 
$\sin(\alpha -\beta) = -0.9999$, 
and (b) $m_{h}^{} =120$ GeV and $m_{H}^{} = m_{A} = m_{H^{\pm}} = 350$ GeV.
The remaining parameter $M$ is scanned from 0 to 1,000 GeV. 
The dark (light) shaded area indicates 
the excluded region by the theoretical requirements of 
vacuum stability and perturbative unitarity 
with $\xi=1$ ($\xi=1/2$).
}
\label{Fig:allowed-kappa}
\end{figure}

In Figs.~\ref{Fig:allowed-kappa}-(a) and \ref{Fig:allowed-kappa}-(b),
contour plots for $|\kappa_{32}^{\text{max}}|^{2}$ are shown 
under the rare tau decay results in the $\tan\beta$-$m_{A}$ plane and
the $\tan\beta$-$\sin^{2}(\alpha-\beta)$ plane, respectively. 
The combined excluded region from the theoretical conditions 
of vacuum stability and pertrbative unitarity 
is indicated by the dark shaded area for 
the criterion $\xi=1$ and by the light one for $\xi=1/2$.
In Fig.~\ref{Fig:allowed-kappa}-(a), 
parameters of the Higgs sector are taken to be
$m_{h}^{} =120$ GeV, $m_{H}^{} = m_{H^{\pm}}^{} = 350$ GeV and 
$\sin(\alpha -\beta) = -0.9999$.
In Fig.~\ref{Fig:allowed-kappa}-(b), those are 
$m_{h}^{} =120$ GeV and $m_{H}^{} = m_{A}^{} = m_{H^{\pm}}^{} = 350$ GeV.
The value of $|\kappa^{\text{max}}_{32}|^{2}$ is independent of 
$M$, the soft-breaking scale of the discrete symmetry. 
On the other hand, theoretical bounds from vacuum stability and 
perturbative unitarity are sensitive to $M$.
Therefore, we evaluate such a theoretical allowed region by scanning 
$M$ to be from 0 to 1000 GeV.
We also take into account the constraint from the $\rho$ parameter 
measurement and the $b\rightarrow s \gamma$ result
by taking $\sin(\alpha-\beta) \simeq -1$ and 
$m_{H}^{}=m_{H^{\pm}}^{}$ with $m_{H^{\pm}}^{}\gtrsim 350$ 
GeV for Fig~\ref{Fig:allowed-kappa}-(a), 
and $m_{A}^{}=m_{H^{\pm}}^{}$ with $m_{H^{\pm}}^{}\gtrsim350$ GeV 
for Fig~\ref{Fig:allowed-kappa}-(b). 
From the both figures, 
it is easily found that
the value of $|\kappa_{32}^{\text{max}}|^{2}$
can extensively be larger for smaller $\tan\beta$
in the allowed region under the theoretical constraints.
For $\tan\beta \lesssim 10$ $(30)$,  $|\kappa_{32}^{\text{max}}|^{2}$ 
can be $\mathcal{O}(0.1)$  ($\mathcal{O}(10^{-4})$).
Among the rare tau decay processes, $\tau \rightarrow \mu\eta$ and
$\tau \rightarrow \mu \gamma$ provide the most stringent
constraints on $|\kappa_{32}^{}|^{2}$.
While $\tau \rightarrow \mu \eta$ is mediated only by $A$,
$\tau \rightarrow \mu\gamma$ depends on the masses of $h$, $H$, 
$A$ and $H^\pm$.
For $\sin^{2}(\alpha-\beta) \sim 1$ and $m_{A}^{} \sim m_{H}^{}$, 
the branching ratio of $\tau\to\mu\gamma$
is suppressed because of the cancellation between the one-loop diagrams 
of $A$ and $H$. 
Therefore, $|\kappa_{32}|^2$ is bounded most strongly 
by the $\tau \rightarrow \mu \eta$ result for this 
case\footnote{The MSSM result approximately corresponds to 
this case\cite{Sher-tmeta}.}.
When $m_A$ differs from $m_H$ or when $\sin^2(\alpha-\beta)$ is to some
extent smaller than unity, the one-loop induced 
$\tau\rightarrow \mu\gamma$ process becomes important, 
and gives the most stringent bound on $|\kappa_{32}|^2$ 
of all the rare tau decay processes.

The value of $|\kappa^{\text{max}}_{32}|^{2}$ can be
much larger than 100 in a wide range of the parameter region. 
One might think that 
such large values of $|\kappa_{3i}|$ cannot be consistent with the
unitarity argument for the LFV Yukawa couplings.
However, 
it should be emphasized that the above figures show 
the contour plots for $|\kappa_{32}^{\text{max}}|^{2}$ under the
rare tau decay results, and not for $|\kappa_{32}|^{2}$.
The region of $|\kappa^{\text{max}}_{32}|^{2} \gtrsim 1$ should be
taken as the area where $|\kappa_{32}|^{2} $ can be  
as large as $\mathcal{O}(10^{-2}\text{-}10^{-4})$ easily. 
It is concluded that current results of the tau LFV decays do not give 
any substantial upper limit on $|\kappa_{32}|^2$ except for 
high $\tan\beta$ region ($\tan\beta \gtrsim 30$). 

Finally, we comment on the bound on $|\kappa_{31}|^2$, 
the LFV parameters for $\tau$-$e$ mixing.
Similar to $\tau$-$\mu$ mixing, we can discuss 
$|\kappa_{31}^{\rm max}|^2$ 
comparing the data of $\tau \to e \eta$, $\tau \to e \mu^+\mu^-$, 
$\tau \to e K^+K^-$ and $\tau \to e \gamma$ listed 
in Table~\ref{Tab:tau-bound} 
with the formulas given in Ref.~\cite{kot}.
These formulas for $\tau$-$e$ mixing 
are common with $\tau$-$\mu$ mixing 
except for the factor of $|\kappa_{3i}|^2$, so that 
difference in contours of $|\kappa_{31}^{\rm max}|^2$ from 
those of $|\kappa_{32}^{\rm max}|^2$ only comes from that in the data. 
In Table~\ref{Tab:tau-bound},
the experimental limit for the branching ratio of $\tau \to e \eta$ is 
about 1.5 times weaker than that of $\tau \to \mu \eta$, while  
that of $\tau \to e \gamma$ is 5.7 times relaxed as compared to
that of $\tau \to \mu \gamma$. 
Moreover, the upper limit on ${\rm Br}(\tau^- \to e^-K^+K^-)$ is 
1.8 times stronger than that on ${\rm Br}(\tau^- \to \mu^-K^+K^-)$.
We have numerically confirmed that 
there are some regions where $\tau^- \to e^-K^+K^-$ can give 
the most stringent bound on $|\kappa_{31}|^2$. 
Therefore, $|\kappa_{31}^{\rm max}|^2$ is determined from one of 
$\tau \to e \eta$, $\tau \to e \gamma$ and $\tau^- \to e^-K^+K^-$ 
depending on parameter regions.

\section{Lepton flavor violating Higgs boson decays}
\label{Sec:LFV-Hdecay}

As shown in the previous section,
the LFV Yukawa couplings can be tested 
only in the large $\tan\beta$ region by searching for rare tau decays. 
In order to cover the region unconstrained by rare tau decay results, 
we here consider LFV via the decay of the neutral Higgs bosons;
i.e., $\phi^{0} \rightarrow \tau^\pm \ell_i^\mp$ ($\phi^{0}=h,H$ and $A$).
Branching ratios for these decays are 
calculated\cite{Rossi,Assamagan,kot,Osaka,Herrero} to be
\begin{eqnarray}
{\rm Br}(h\rightarrow \tau^{-} \ell^{+}_i) =&
 \frac{1}{16\pi} 
 \frac{m_{\tau}^{2} \cos^{2} \left(\alpha - \beta \right)}
 {v^{2} \cos^{4}\beta}
 \left| \kappa_{3i} \right|^{2}
 \frac{m_{h}^{} \left(1- \frac{m_{\tau}^{2}}{m_{h}^{2}}\right)^{2}}
 {\Gamma(h\rightarrow\text{all})}, 
\label{eq:Brhtm}\\
{\rm Br}(H\rightarrow \tau^{-} \ell^{+}_i) =&
 \frac{1}{16\pi} 
 \frac{m_{\tau}^{2} \sin^{2} \left(\alpha - \beta \right)}
 {v^{2} \cos^{4}\beta}
 \left| \kappa_{3i} \right|^{2}
 \frac{m_{H}^{} \left(1- \frac{m_{\tau}^{2}}{m_{H}^{2}}\right)^{2}}
 {\Gamma(H\rightarrow\text{all})},
\label{eq:BrhtmH}\\
{\rm Br}(A\rightarrow \tau^{-} \ell^{+}_i) =&
 \frac{1}{16\pi} 
 \frac{m_{\tau}^{2}}
 {v^{2} \cos^{4}\beta}
 \left| \kappa_{3i} \right|^{2}
 \frac{m_{A}^{} \left(1- \frac{m_{\tau}^{2}}{m_{A}^{2}}\right)^{2}}
 {\Gamma(A\rightarrow\text{all})},
\label{eq:BrhtmA}\end{eqnarray}
where $\Gamma (\phi^{0} \rightarrow \text{all}) $  
is the total width for corresponding neutral Higgs boson $\phi^{0}$.
We here neglect terms of $\mathcal{O}(m_{\ell_i}^{2}/m_{\phi^{0}}^{2})$.
Branching ratios for $\phi^0 \to \tau^+\ell_i^-$ coincide with 
those for $\phi^0 \to \tau^-\ell_i^+$ given in 
Eqs.~(\ref{eq:Brhtm}), (\ref{eq:BrhtmH}) and (\ref{eq:BrhtmA}).
In the following, we concentrate on the decays into a $\tau$-$\mu$
pair. We take the values of the SM parameters as 
$\alpha_{\text{em}}=0.007297$, $G_{F}=1.166 \times 10^{-5}$
$\text{GeV}^{-2}$,
$m_{Z}^{}=91.19$ GeV,
$m_{\tau} = 1.777$ GeV, $m_{\mu}=0.1057$ GeV, 
$m_{b}=4.1$ GeV, $m_{t}=174.3$ GeV, $m_{c}=1.15$ GeV, $m_{s}=0.120$ GeV.


A search for the LFV decays $h\rightarrow \tau^\pm\ell_i^\mp$ 
can give important information for extended Higgs sectors and thus 
for the structure of new physics,
even when only $h$ is found and any other direct signals for the extended
Higgs sector are not obtained by experiments.
We here evaluate the possible maximal value 
of the branching ratio 
${\rm Br}(h \rightarrow \tau^{-} \mu^{+})_{\text{max}}$
under the results of the rare tau decay search,
by inserting $|\kappa_{32}^{\text{max}}|^{2}$ 
of Eq.~(\ref{eq:kappaSq-max}) into
the $|\kappa_{32}|^{2}$ in Eq.~(\ref{eq:Brhtm}).

\begin{figure}[t]
\begin{minipage}{8cm}
\unitlength=1cm
\begin{picture}(7.8,6)
\put(0,1){\includegraphics[width=8cm]{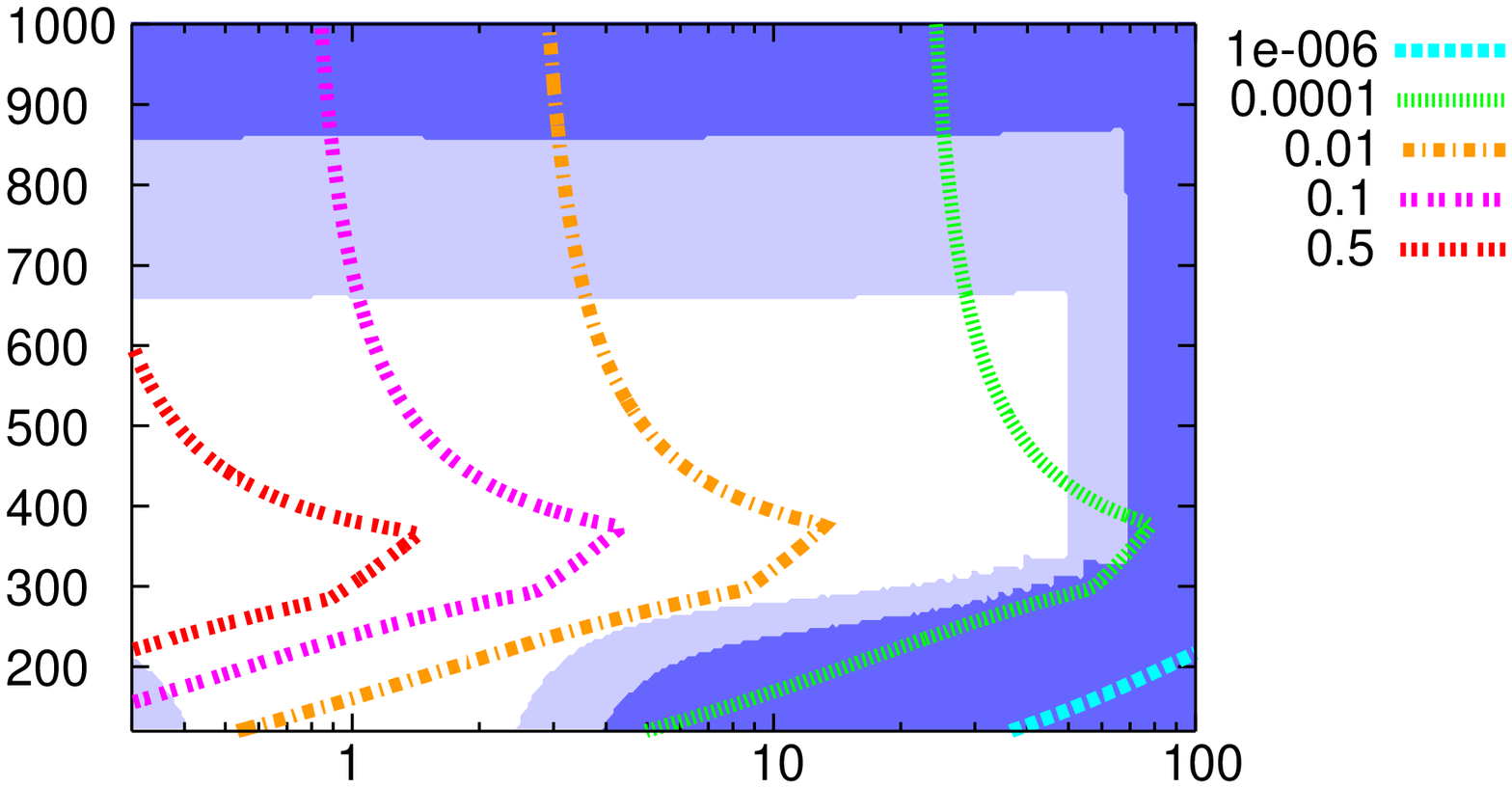}}
\put(3,0.5){(a)}
\put(0,5.5){$m_{A}^{}$ [GeV]}
\put(6.8,1){$\tan\beta$}
\end{picture}
\end{minipage}
\begin{minipage}{8cm}
\unitlength=1cm
\begin{picture}(7.8,6)
\put(0,1){\includegraphics[width=7.8cm]{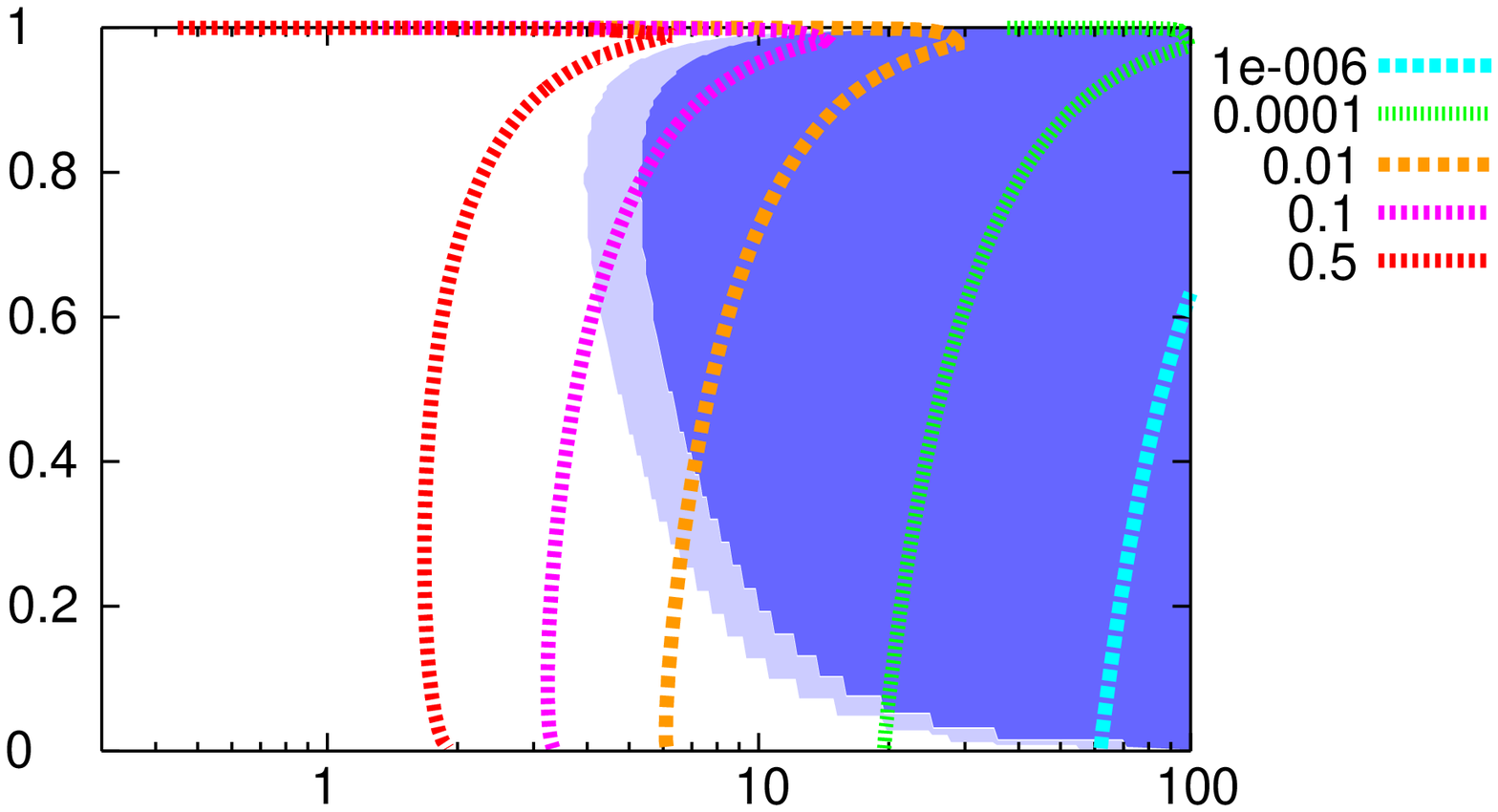}}
\put(3,0.5){(b)}
\put(0,5.5){$\sin^{2}(\alpha-\beta)$}
\put(6.8,1){$\tan\beta$}
\end{picture}
\end{minipage}
\caption{Contour plots of 
 ${\rm Br}(h \rightarrow \tau^{\pm} \mu^{\mp})_{\text{max}}$, 
 the possible maximal values for the branching ratio 
 under the tau rare decay results, are shown (a) in  
 the $\tan\beta$-$m_{A}$ plane  
 and (b) in the $\tan\beta$-$\sin(\alpha-\beta)$ plane. 
 The parameters are taken as the same as 
 Figs.~\ref{Fig:allowed-kappa}-(a) and \ref{Fig:allowed-kappa}-(b), 
 respectively. 
 The dark (light) shaded area indicates 
 the excluded region by the theoretical requirements of 
vacuum stability and perturbative unitarity with $\xi=1$ 
($\xi=1/2$).
}
\label{Fig:Br-htm}
\end{figure}

In Figs.~\ref{Fig:Br-htm}-(a) and \ref{Fig:Br-htm}-(b), contours of 
${\rm Br}(h \rightarrow \tau^{\pm}\mu^{\mp})_{\text{max}}$, which is 
twice of ${\rm Br}(h \rightarrow \tau^{-}\mu^{+})_{\text{max}}$, 
are shown in the $\tan\beta$-$m_{A}^{}$ plane and
in the $\tan\beta$-$\sin^{2}(\alpha-\beta)$ plane, respectively.
The parameters are taken to be the same as those for
Figs.~\ref{Fig:allowed-kappa}-(a) and \ref{Fig:allowed-kappa}-(b), 
respectively; i.e.,  
(a) $m_{h}^{} =120$ GeV, $m_{H}^{} = m_{H^{\pm}}^{} = 350$ GeV and 
    $\sin(\alpha -\beta) = -0.9999$, and 
(b) $m_{h}^{} =120$ GeV and $m_{H}^{} = m_{A}^{} = m_{H^{\pm}}^{} = 350$
GeV, with $M$ to be scanned from 0 to 1000 GeV.
We again show the excluded area from requirements of tree-level
unitarity and vacuum stability as in the same way 
as Figs.~\ref{Fig:allowed-kappa}-(a) and \ref{Fig:allowed-kappa}-(b). 
For low and moderate values of $\tan\beta$  ($\tan\beta \lesssim 30$),  
where rare tau decay results cannot give
substantial upper limit on $|\kappa_{32}|^{2}$,
${\rm Br}(h \rightarrow \tau^{\pm}\mu^{\mp})_{\text{max}}$ 
can be sufficiently large. 
We find that the possible maximal values of the 
branching ratio can be greater than $\mathcal{O}(10^{-3})$ 
in a wide rage of the theoretically allowed region.

For relatively lower $\tan\beta$ values,
the experimental upper limits on $|\kappa_{32}|^2$ 
from rare tau decays are weaker, and 
$\text{Br}(h\rightarrow\tau^{\pm} \mu^{\mp})$ 
can be sufficiently large 
($\gtrsim \mathcal{O}(10^{-3})$ for $m_h \sim 120$ GeV).      
It is expected that a sufficient number of 
such light $h$ can be produced at future 
colliders such as CERN LHC,  
currently planned International Linear Collider (ILC) and 
CERN CLIC. 
It has been pointed out that the decay process $h \to \tau^\pm\mu^\mp$
can easily be detected at ILC with the luminosity of 1 ab$^{-1}$,  
when $m_h \sim 120$ GeV and 
${\rm Br}(h\rightarrow\tau^\pm\mu^\mp) \gtrsim \mathcal{O}(10^{-3})$
via the Higgsstrahlung process by using the recoil 
momentum of $Z$ boson\cite{Osaka}. 
Therefore, the LFV search via the decay $h \to \tau^\pm\mu^\mp$ 
at ILC is complementary to that via rare tau decays 
at (super) B factories, and the both 
cover a wide region of the parameter space 
of the lepton flavor violating THDM.


Next we discuss branching ratios for the LFV decays of 
heavier Higgs bosons, $H/A \to \tau^\pm \ell_i^\mp$, 
using Eqs.~(\ref{eq:BrhtmH}) and (\ref{eq:BrhtmA}) under the 
current data of LFV rare tau decays. 
In the THDM, there are many possible decay 
modes for $H$ depending on the mass spectrum. 
In the numerical analysis, we included contributions from 
one-loop induced $Z\gamma$, $\gamma\gamma$ and $gg$ modes 
in addition to all the tree level modes.
Those for $A$ are one-loop induced modes of 
$hh$, $hH$, $HH$, $h\gamma$, $H\gamma$, 
$W^\pm W^\mp$, $ZZ$, $Z\gamma$, $\gamma\gamma$ and $gg$ 
in addition to all the tree level modes.
The branching ratios for $H/A \to \tau^\pm \ell_i^\mp$ 
are sensitive to the masses of all the Higgs bosons.
Here we consider the case of 
$\sin(\alpha-\beta)=-1$ and  
$m_H^{}=m_A^{}=m_{H^\pm}^{}$ $(\equiv m_\Phi^{})$. 
As discussed in Sec. II, 
the $\rho$ parameter constraint is satisfied 
for this choice. From the $b \to s \gamma$
results, $m_\Phi^{}$ is taken to be greater than $350$ GeV.  
As also discussed in Sec. II, $M$ 
determines the decoupling property of heavier Higgs bosons. 
Although the branching ratios ${\rm Br}(H/A \to \tau^\pm \ell_i^\mp)$ 
are insensitive to $M$ in the present parameter set, its value 
strongly affects the allowed parameter region under the theoretical 
conditions of vacuum stability and perturbative unitarity.
Notice that couplings of $H$ are similar to those of $A$ for 
$\sin(\alpha-\beta)=-1$ where there are no $HVV$ couplings. 
Hence we show the results only for the LFV decays of $H$ below. 
In a general case, the branching ratio of $H \to \tau^\pm \mu^\mp$ 
tends to be smaller than 
that of $A \to \tau^\pm \mu^\mp$ due to the contribution from 
the modes $H \to VV$.

\begin{figure}[t]
\begin{minipage}{5cm}
\unitlength=1cm
\begin{picture}(7.8,7)
\put(0,1){\includegraphics[width=5.14cm]{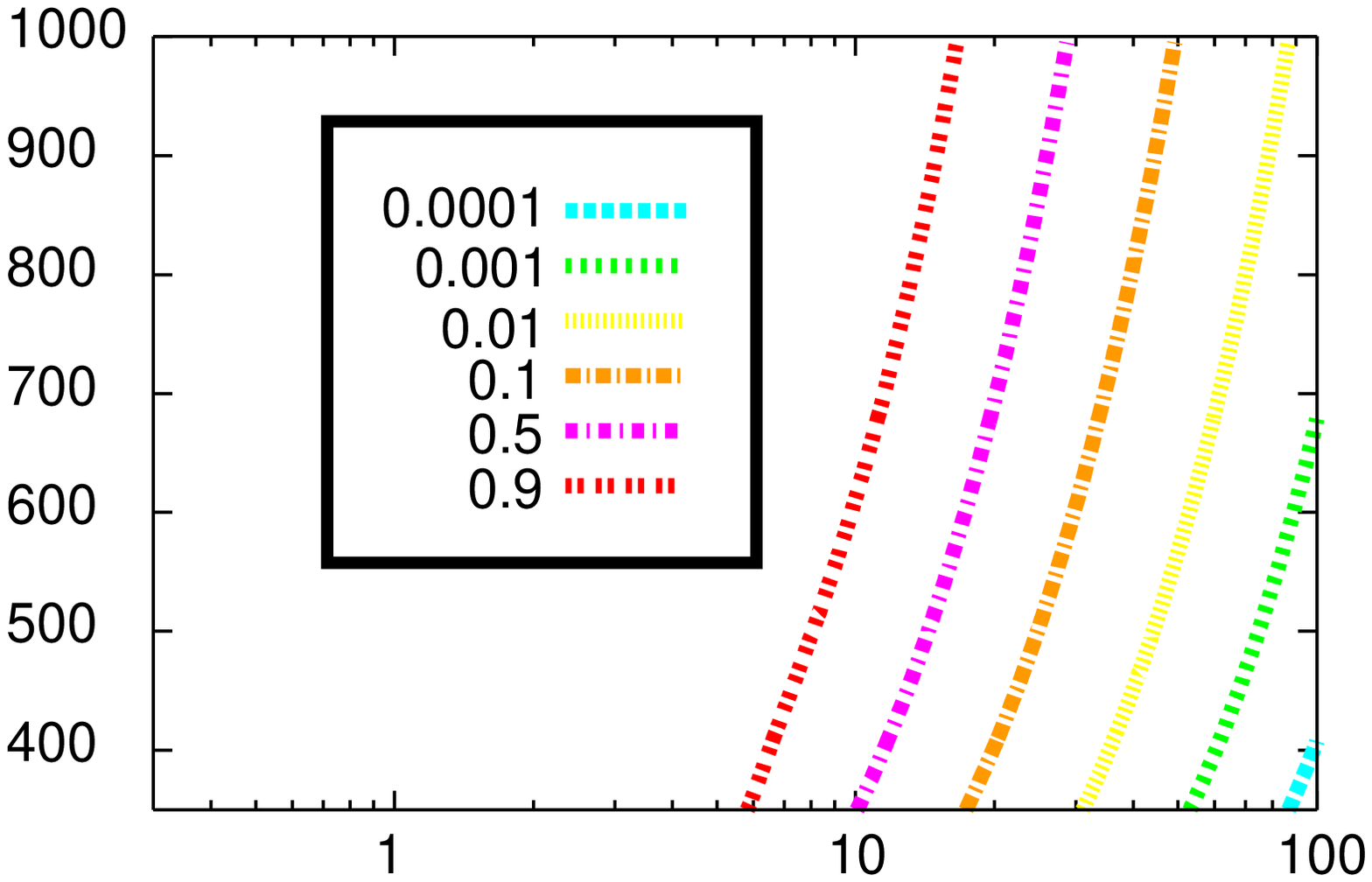}}
\put(2,0.5){(a)}
\put(0,4.6){$m_{\Phi}^{}$ [GeV]}
\put(4,0.7){$\tan\beta$}
\end{picture}
\end{minipage}
\begin{minipage}{5cm}
\unitlength=1cm
\begin{picture}(7.8,7)
\put(0.24,1){\includegraphics[width=4.58cm]{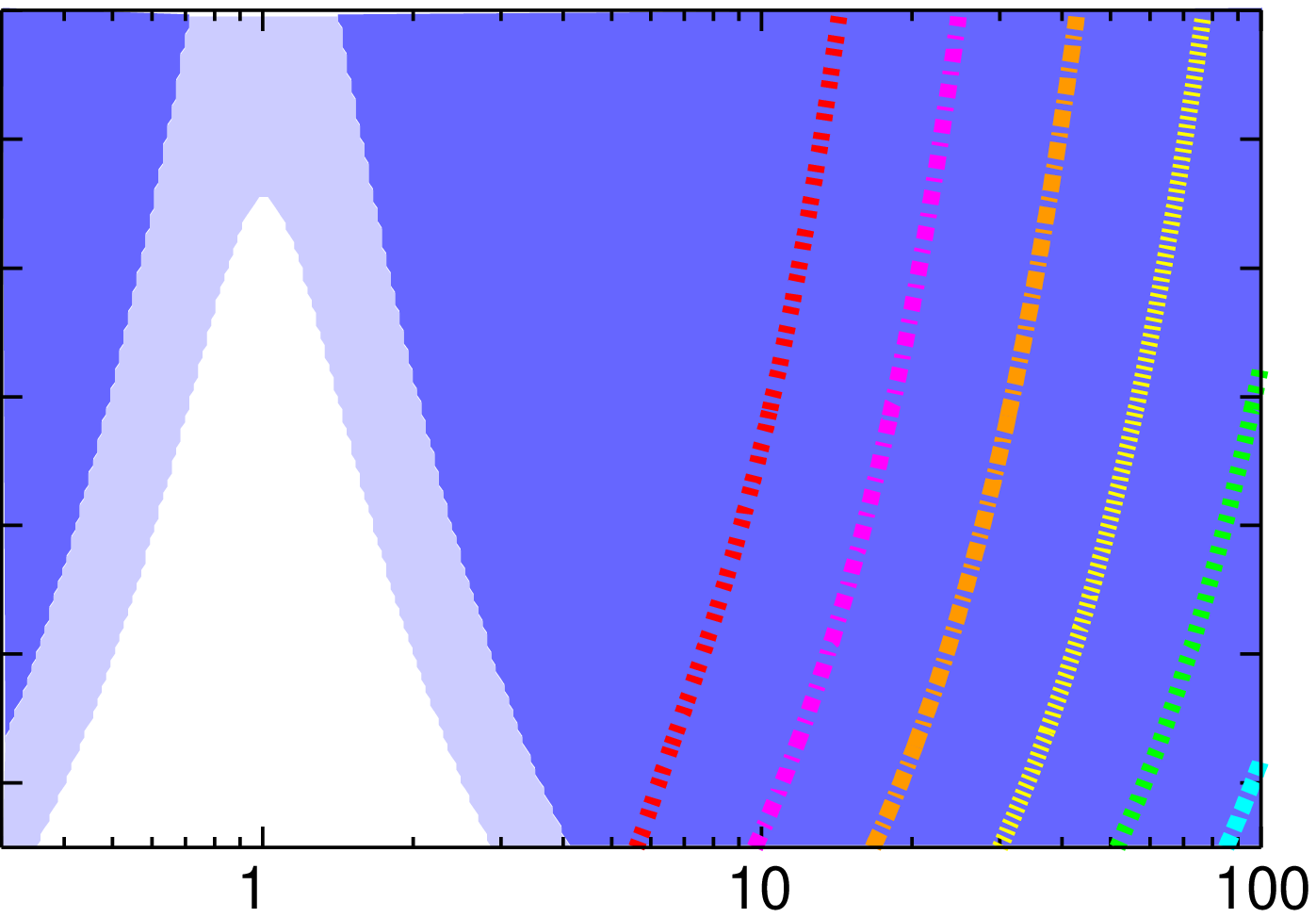}}
\put(2,0.5){(b)}
\put(4,0.7){$\tan\beta$}
\end{picture}
\end{minipage}
\begin{minipage}{5cm}
\unitlength=1cm
\begin{picture}(7.8,7)
\put(0,1){\includegraphics[width=4.52cm]{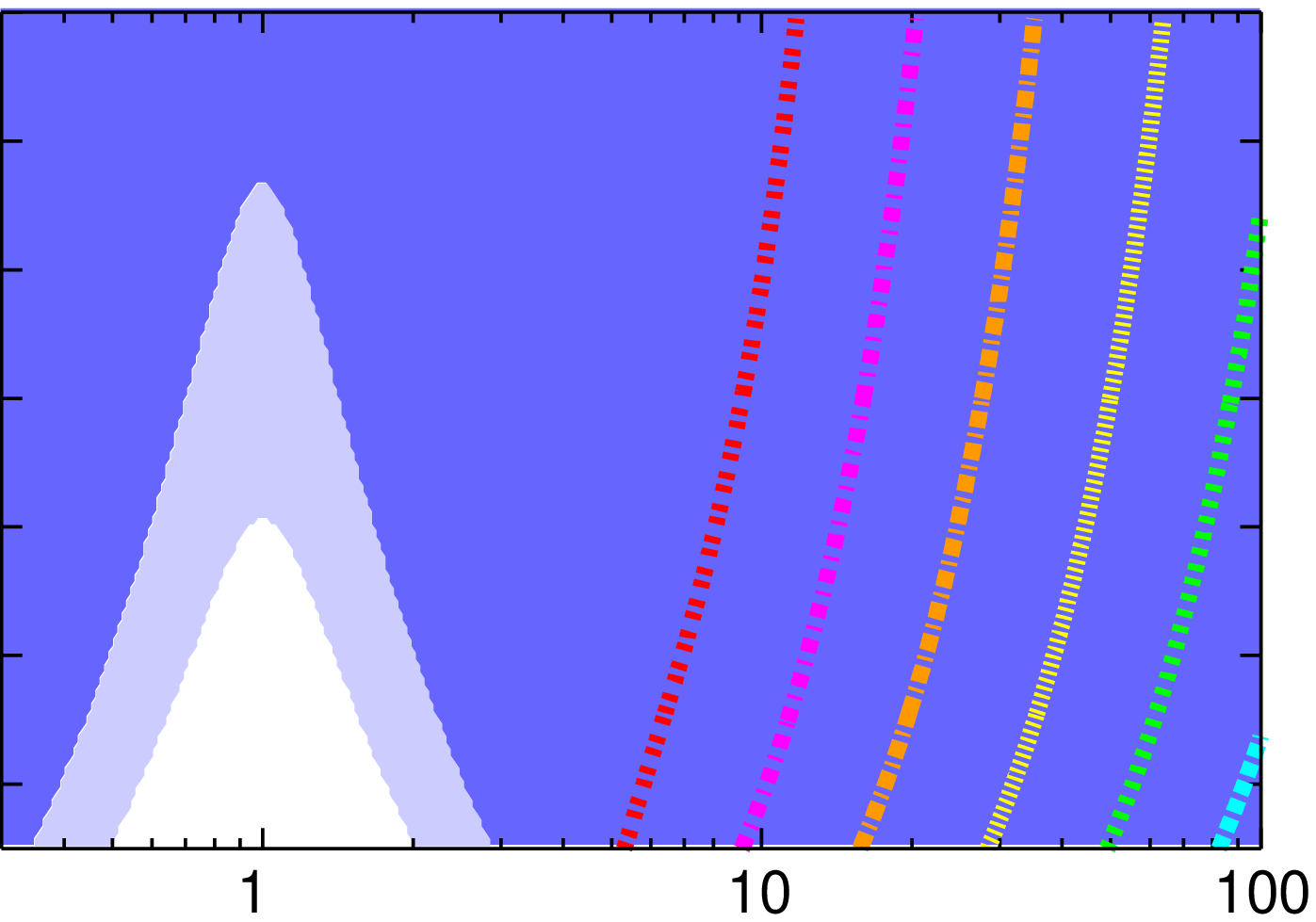}}
\put(2,0.5){(c)}
\put(4,0.7){$\tan\beta$}
\end{picture}
\end{minipage}
\caption{
Contour plots of 
${\rm Br}(H \rightarrow \tau^{\pm} \mu^{\mp})_{\text{max}}$ 
are shown in the $\tan\beta$-$m_{\Phi}^{}$ plane 
($m_\Phi^{}\equiv m_H^{}=m_A^{}=m_{H^\pm}^{}$)
for $m_h^{}=120$ GeV and $\sin(\alpha-\beta)=-1$ with 
(a) $M=m_\Phi^{}$,  
(b) $M=m_\Phi^{}/\sqrt{2}$, and  
(c) $M=0$. 
The dark (light) shaded area indicates 
the excluded region by the theoretical requirements of 
vacuum stability and perturbative unitarity 
with $\xi=1$ ($\xi=1/2$).
}
\label{Fig:Br-Htm}
\end{figure}

In Figs.~\ref{Fig:Br-Htm}-(a), 
\ref{Fig:Br-Htm}-(b) and \ref{Fig:Br-Htm}-(c), 
contour plots of ${\rm Br}(H \to \tau^\pm\mu^\mp)_{\rm max}$, 
the upper limit of ${\rm Br}(H \to \tau^\pm\mu^\mp)$ under the rare 
tau decay results, are shown in the $\tan\beta$-$m_\Phi^{}$ 
plane for $M=m_\Phi^{}$, $m_\Phi^{}/\sqrt{2}$ and $0$, respectively.
As expected, the contours are insensitive to the values of 
$M$, and approximately the same in 
Figs.~\ref{Fig:Br-Htm}-(a), \ref{Fig:Br-Htm}-(b) and \ref{Fig:Br-Htm}-(c). 
It is shown that ${\rm Br}(H \to \tau^\pm\mu^\mp)_{\rm max}$ 
can be larger than $10^{-3}$ except for large 
$\tan\beta$ values with relatively small $m_\Phi^{}$. 
Therefore, it turns out to be no substantial upper 
limit on the ${\rm Br}(H \to \tau^\pm\mu^\mp)$ in the 
relatively low $\tan\beta$ region ($\tan\beta \lesssim 20$)
from the LFV rare tau decay results. 
When $M$ is smaller than $m_\Phi$, where 
the heavier Higgs boson partially receive their masses from 
the vacuum expectation value, the allowed parameter region 
is strongly constrained by the requirements 
of vacuum stability and perturbative unitarity.
In particular, for $M=0$ (Fig.~\ref{Fig:Br-Htm}-(c)), 
the allowed region is limited only the area 
of around $\tan\beta \sim 1$ and $m_\Phi^{} \lesssim 600$ GeV. 

The extra Higgs bosons ($H$, $A$ and $H^\pm$) 
are expected to be searched at the LHC. 
The signal of $gg \to H/A \to \tau^\pm \mu^\mp$ 
may be detectable at LHC with high luminosity (100 ${\rm fb}^{-1}$) 
when ${\rm Br}(H/A \to \tau^\pm\mu^\mp)$ is 
greater than $10^{-2}$ for $m_{H/A}^{} \sim 350$ GeV and 
$\tan\beta=45$\cite{Assamagan}. 
However the rate is rapidly reduced for smaller values of $\tan\beta$ 
and for larger values of $m_{H/A}^{}$. 
Further feasibility study is necessary.

\section{Search for LFV decays of Higgs bosons at a Linear Collider} 

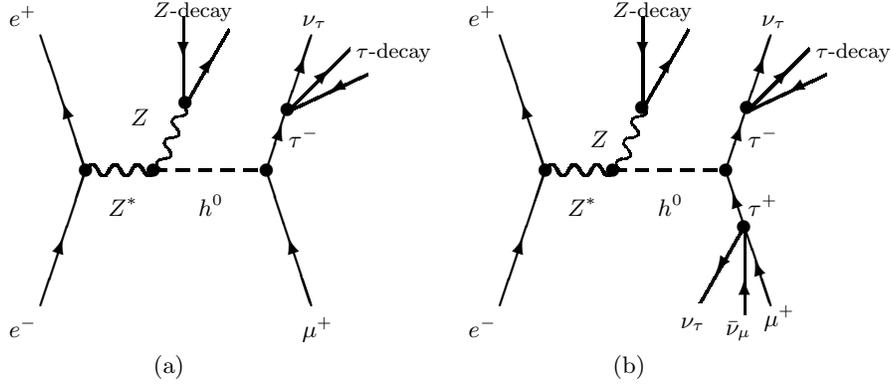
\begin{figure}[t]
\unitlength=0.6cm
\begin{picture}(10,8)
\thicklines
 \put(1,1){\line(1,3){1}}
 \put(1,7){\line(1,-3){1}}
 \multiput(2,4)(0.5,0){3}{
  \qbezier(0,0)(0.125,0.25)(0.25,0)
  \qbezier(0.25,0)(0.375,-0.25)(0.5,0)}
 \put(3.5,4){\rotatebox{70}{
   \multiput(0,0)(0.5,0){3}{
   \qbezier(0,0)(0.125,0.25)(0.25,0)
   \qbezier(0.25,0)(0.375,-0.25)(0.5,0)}}}
  \put(4.15,5.4){\rotatebox{90}{\line(1,0){2}}}
  \put(4.15,5.4){\rotatebox{60}{\line(1,0){2}}}
 \multiput(3.5,4)(0.5,0){5}{\line(1,0){0.3}}
 \put(6,4){\line(1,3){1}}
 \put(6,4){\line(1,-3){1}}
  \put(6.42,5.3){\rotatebox{45}{\line(1,0){2}}}
  \put(6.42,5.3){\rotatebox{25}{\line(1,0){2}}}
 \put(2,4){\circle*{0.3}}
 \put(3.5,4){\circle*{0.3}}
 \put(6,4){\circle*{0.3}}
 \put(4.2,5.5){\circle*{0.3}}
 \put(6.47,5.35){\circle*{0.3}}
 \put(1.5,2.5){\vector(1,3){0}}
 \put(1.505,5.5){\vector(-1,3){0}}
 \put(6.5,2.5){\vector(-1,3){0}}
 \put(6.35,5){\vector(1,3){0}}
 \put(4.16,6.5){\vector(0,-1){0}}
 \put(4.82,6.5){\vector(2,3){0}}
 \put(6.85,6.5){\vector(1,3){0}}
 \put(7.4,6.25){\vector(1,1){0}}
 \put(7.4,5.75){\vector(-2,-1){0}}
 \put(0.3,0.3){$e^{-}$} \put(0.3,7.2){$e^{+}$}
 \put(2.5,3){$Z^{*}$} \put(3,5){$Z$}
 \put(4.5,3){$h^{0}$}
 \put(6.8,0.3){$\mu^{+}$} 
 \put(6.5,4.5){$\tau^{-}$} 
 \put(6.8,7.2){$\nu_{\tau}$}
 \put(8,6.5){\footnotesize $\tau$-decay}
 \put(3.5,7.4){\footnotesize $Z$-decay}
 \put(3.5,-0.5){(a)}
\end{picture}
\unitlength=0.6cm
\begin{picture}(10,8)
\thicklines
 \put(1,1){\line(1,3){1}}
 \put(1,7){\line(1,-3){1}}
 \multiput(2,4)(0.5,0){3}{
  \qbezier(0,0)(0.125,0.25)(0.25,0)
  \qbezier(0.25,0)(0.375,-0.25)(0.5,0)}
 \put(3.5,4){\rotatebox{70}{
   \multiput(0,0)(0.5,0){3}{
   \qbezier(0,0)(0.125,0.25)(0.25,0)
   \qbezier(0.25,0)(0.375,-0.25)(0.5,0)}}}
  \put(4.15,5.4){\rotatebox{90}{\line(1,0){2}}}
  \put(4.15,5.4){\rotatebox{60}{\line(1,0){2}}}
 \multiput(3.5,4)(0.5,0){5}{\line(1,0){0.3}}
 \put(6,4){\line(1,3){1}}
 \put(6,4){\line(1,-3){1}}
  \put(6.42,5.3){\rotatebox{45}{\line(1,0){2}}}
  \put(6.42,5.3){\rotatebox{25}{\line(1,0){2}}}
  \put(6.4,2.8){\rotatebox{-90}{\line(1,0){2}}}
  \put(6.4,2.8){\rotatebox{60}{\line(-1,0){2}}}
 \put(2,4){\circle*{0.3}}
 \put(3.5,4){\circle*{0.3}}
 \put(6,4){\circle*{0.3}}
 \put(4.15,5.4){\circle*{0.3}}
 \put(6.4,2.75){\circle*{0.3}}
 \put(6.47,5.4){\circle*{0.3}}
 \put(1.5,2.5){\vector(1,3){0}}
 \put(1.505,5.5){\vector(-1,3){0}}
 \put(6.15,3.5){\vector(-1,3){0}}
 \put(6.35,5){\vector(1,3){0}}
 \put(4.16,6.5){\vector(0,-1){0}}
 \put(4.82,6.5){\vector(2,3){0}}
 \put(6.85,6.5){\vector(1,3){0}}
 \put(7.4,6.25){\vector(1,1){0}}
 \put(7.4,5.75){\vector(-2,-1){0}}
 \put(6.655,2){\vector(-1,3){0}}
 \put(6.41,1.6){\vector(0,1){0}}
 \put(5.8,1.745){\vector(-2,-3){0}}
 \put(0.3,0.3){$e^{-}$} \put(0.3,7.2){$e^{+}$}
 \put(2.5,3){$Z^{*}$} 
 \put(3,4.5){$Z$} 
 \put(4.5,3){$h^{0}$}
 \put(6.8,7.2){$\nu_{\tau}$}
 \put(6.5,4.5){$\tau^{-}$}
 \put(6.5,3){$\tau^{+}$}
 \put(6.85,0.55){$\mu^{+}$}
 \put(6,0.35){$\bar{\nu}_{\mu}$} 
 \put(5,0.5){$\nu_{\tau}$}
 \put(8,6.5){\footnotesize $\tau$-decay}
 \put(3.5,7.4){\footnotesize $Z$-decay}
  \put(3.5,-0.5){(b)}
\end{picture}
\caption{The Feynman diagram of the signal event (a), and that of {\it the fake event} (b).}
\label{Fig:diagrams}
\end{figure}

Let us consider the LFV  Higgs decay 
$h^{0} \to \tau^{\pm} \mu^{\mp}$ at a LC 
in the situation where the heavier Higgs bosons 
nearly
decouple from the gauge bosons; i.e.,
$\sin(\alpha-\beta) \simeq -1$. 
The lightest Higgs boson then approximately behaves as the SM one.  
The main production modes of the lightest Higgs boson at a LC
are the Higgsstrahlung $e^+e^- \to Z^\ast \to Z h^{0}$ and the $W$ fusion 
$e^+e^- \to (W^{+\ast} \bar{\nu}_{e})(W^{-\ast} \nu_{e}) 
\to h^{0} \nu_{e} \bar{\nu}_{e}$. 
For a light $h^0$ with the mass $m_h \sim 120$ GeV, 
the former production mechanism is dominant at low collision 
energies ($\sqrt{s} < 400$-$500$ GeV), while the latter dominates 
at higher energies. 
For our purpose, the Higgsstrahlung process 
is useful because of its simple kinematic structure. 
The signal process is then 
$e^+e^- \to Z^\ast \to Z h^{0} \to Z \tau^\pm \mu^\mp$.
We can detect the outgoing muon with high efficiency, 
and its momentum can be measured precisely by event-by-event. 
The momentum of the $Z$ boson can be reconstructed from those of 
its leptonic $\ell^+\ell^-$ 
($\ell^\pm = e^\pm$ and $\mu^\pm$) or hadronic ($jj$) decay products. 
Therefore, we can identify the signal event
without measuring $\tau$ momentum directly, 
as long as the beam spread rate for $\sqrt{s}$ 
is sufficiently low.

Depending on the $Z$ decay channel, the signal events are separated 
into two categories, 
$jj\tau^{\pm}\mu^{\mp}$ and $\ell^+\ell^-\tau^{\pm}\mu^{\mp}$.  
The energy resolution of the $Z$ boson 
from hadronic jets $jj$ is expected to
be $0.3\sqrt{E_{Z}}$ GeV   
and that from $\ell^+\ell^-$ is $0.1 \sqrt{E_{Z}}$ GeV.
We assume that the detection efficiencies 
of the $Z$ boson and the muon are 100 \%,   
the rate of the beam energy spread is expected to be 0.1 \% level, 
the muon momentum is measured with high precision 
and the mass of the lightest Higgs boson will have been determined 
in the 50 MeV level. We also expect that the 
effect of the initial state radiation is small for the collider 
energies that we consider ($\sqrt{s} \sim 250$-$300$ GeV).
Taking into account all these numbers, we expect that the tau momentum 
can be determined indirectly within 3 GeV for 
$jj\tau^{\pm}\mu^{\mp}$ and 1 GeV for $\ell^+\ell^- \tau^{\pm}\mu^{\mp}$. 

Let us evaluate the number of the signal event. 
We assume that the energy $\sqrt{s}$ is tuned depending 
on the mass of the lightest Higgs boson: i.e.,
we take the optimal $\sqrt{s}$ to product the lightest Higgs boson
through the Higgsstrahlung process.
(It is approximately given by 
$\sqrt{s} \sim m_{Z} + \sqrt{2} m_{h}^{}$.)
The production cross section of $e^+e^- \to Z h^{0}$ is about 
$220$ fb for $m_{h}^{}=123$ GeV.
Then, we obtain $2.2 \times 10^5$ Higgs events 
if the integrated luminosity is 1 ab$^{-1}$.
When $|\kappa_{32}|^2 $ is 
$8.4 \times 10^{-6}$, 
about $118$ events of $jj \tau^\pm\mu^\mp$ and 
$11$ events of $\ell^+\ell^-\tau^{\pm}\mu^{\mp}$ 
can be produced. 
 
Next, we consider the background.  
For the signal with the Higgs boson mass of $\sim 120$ GeV, 
the main background comes from $e^+e^- \to Z \tau^+\tau^- $. 
The number of the $Z \tau^{\pm}\mu^{\mp}$ event from 
$e^+e^- \to Z \tau^+\tau^-$ is estimated 
about $3.6 \times 10^{4}$.
Although the number of the background events is huge,
we can expect that a large part of them is effectively 
suppressed by using the kinematic cuts\cite{Osaka}.
The irreducible background comes from the process shown in 
Fig.\ref{Fig:diagrams}-(b): 
the Higgs boson decays into a
tau pair, and one of the tau decays into a muon and missings ($e^{+}
e^{-} \rightarrow Z h^{0} \rightarrow Z \tau^{+}\tau^{-} \rightarrow
Z \tau^{\pm}\mu^{\mp}$+missings).
We can not distinguish the signal event $h^{0}\rightarrow \tau^{\pm}\mu^{\mp}$
with the event of Fig.\ref{Fig:diagrams}-(b) when the muon emitted 
from the tau lepton carries 
the similar momentum to that of the parent, 
because it leaves the same track on the detector as the signal event.
We refer this kind of the background as {\it the fake signal}. 
As examined in Ref.~\cite{Osaka}, 
the number of the fake signal strongly 
depends on the precision of the 
tau momentum determination by the recoil method.
We expect that it is attained with the similar
precision to that of the Higgs boson mass 
reconstructed by the recoil momentum.
We here take the uncertainty of the tau momentum as 3 GeV
for $jj\tau^{\pm}\mu^{\mp}$ and as 1 GeV for 
$\ell^+\ell^- \tau^{\pm}\mu^{\mp}$. 

Finally, we estimate the statistical significance ($S/\sqrt{B}$) 
for each channel. The number of the fake events is evaluated 
in Ref.~\cite{Osaka}, which is 
460 for $jj\tau^{\pm}\mu^{\mp}$ and 15 for 
$\ell^+\ell^- \tau^{\pm} \mu^{\mp}$.
Therefore, when $|\kappa_{32}|^2 $ is $8.4 \times 10^{-6}$ 
with $m_h^{}=123$ GeV, the significance can become 5.5 and 3.0 
for $jj \tau^\pm\mu^\mp$ and $\ell^+\ell^-\tau^{\pm}\mu^{\mp}$, 
respectively, 
taking into account the constraint from the
rare tau decay results.
The combined significance can reach to 6.3. 
When $|\kappa_{32}|^2 $ is $3.8 \times 10^{-6}$ 
with $m_h^{}=123$ GeV,  
the number of the signal becomes smaller, and the combined 
significance amounts to be as large as 2.0.

\section{Summary}

Lepton flavor violating decays of Higgs bosons 
have been studied in the general THDM, 
in which LFV couplings are introduced as a deviation from 
Model II Yukawa interaction in the lepton sector.
The model parameters are constrained by 
requirements of tree-level unitarity and vacuum stability, 
and also from the experimental results. 
The parameters $|\kappa_{3i}|^2$ in LFV Yukawa interactions  
are bounded from above by using the current data for rare tau LFV decays. 
In the large $\tan\beta$ region ($\tan\beta \gtrsim 30 $),  
the upper limit  on $|\kappa_{3i}|^2$ due to the rare tau 
decay data turns out to be 
substantial and comparable with the value predicted 
by assuming some fundamental theories such as SUSY. 
For smaller values of $\tan\beta$, the upper limit
is rapidly relaxed, and no more substantial constraint is obtained from 
the rare tau decay results. 

We have shown that a search for the LFV decays 
$\phi^{0} \rightarrow \tau^\pm \ell_i^\mp$ of neutral Higgs bosons 
($\phi^{0} =h,H$ and $A$) can be useful 
to further constrain the LFV Yukawa couplings
at future collider experiments. 
In particular, the decays of the lightest Higgs boson 
can be one of the important probes to find the evidence for the 
extended Higgs sector even when the SM-like situation would 
be preferred by the data at forthcoming collider experiments.
The branching ratio for $h \to \tau^\pm \mu^\mp$ can be 
larger than $\mathcal{O}(10^{-3})$ 
except for the high $\tan\beta$ region
under the constraints from the current experimental data 
and also from the theoretical requirements. 
At ILC (and in case at LHC), 
these branching fractions can be tested. 
Therefore, we conclude that the search of LFV in the Higgs 
boson decay at future colliders 
can further constrain the LFV Yukawa couplings 
in the parameter region 
where rare tau decay data cannot reach.

Note added: Recently similar work was done 
in Ref.~\cite{paradisi} on the experimental upper bound on 
$|\kappa_{3i}|^2$.


\begin{thebibliography}{9}   



\bibitem{KunoOkada}
        Y.~Kuno and Y.~Okada, {\em Rev. Mod. Phys.} {\bf 73}, {151} 
{(2001)}.
        
\bibitem{belle-tau-lp0} 
        Y.~Enari {\it et al.}, the Belle Collaboration, 
        arXiv:hep-ex/0503041.

\bibitem{belle-tau-lMM} 
       Y.~Yusa, 
       {{\em Nucl. Phys.} B} {\rm (Proc. Suppl.)} 
       {\bf 144}, {173} {(2005)}.      

\bibitem{babar-tau-lMM} 
       M.~Hodkinson (on behalf of the BaBar Collaboration), 
       {{\em Nucl. Phys.} B} {\rm (Proc. Suppl.)} 
       {\bf 144}, {167} {(2005)}.      

\bibitem{belle-tau-3l} 
        Y.~Yusa {\it et al.}, the Belle Collaboration, 
        \Journal{\PLB}{589}{103}{2004}.

\bibitem{babar-tau-3l} 
        B.~Aubert {\it et al.}, the BaBar Collaboration, 
        \Journal{\PRL}{92}{121801}{2004}.

\bibitem{belle-tau-egamma} 
        K.~Hayasaka {\it et al.}, the Belle Collaboration,
        \Journal{\PLB}{613}{20}{2005}.

\bibitem{belle-tau-mugamma} 
        K.~Abe {\it et al.}, the Belle Collaboration,
        \Journal{\PRL}{92}{171802}{2004}.

\bibitem{babar-tau-mugamma}   
        B.~Aubert {\it et al.}, the BaBar Collaboration, 
        arXiv:hep-ex/0502032.

\bibitem{KitanoKoikeKomineOkada}
        R.~Kitano, M.~Koike, S.~Komine, and Y.~Okada, 
        \Journal{\PLB}{575}{300}{2003}.

\bibitem{muegamma-type3THDM} 
        R.A.~Diaz, R.~Martinez, and J.-A. Rodriguez, 
        \Journal{\PRD}{63}{096007}{2001}.

\bibitem{BK}
        K.S.~Babu and C.~Kolda,
        \Journal{\PRL}{89}{241802}{2002}.

\bibitem{DER}
        A.~Dedes, J.~Ellis, and M.~Raidal,
        \Journal{\PLB}{549}{159}{2002}.

\bibitem{Sher-tmeta}
        M.~Sher,
        \Journal{\PRD}{66}{057301}{2002}.

\bibitem{Rossi}
        A.~Brignole and A.~Rossi,
        \Journal{\PLB}{566}{217}{2003}; 
        \Journal{\NPB}{701}{3}{2004}.

\bibitem{cheng-sher}
        T.P.~Cheng and M.~Sher, \Journal{\PRD}{35}{3484}{1987}.

\bibitem{HiggsLFV-THDM}
        J.L.~Diaz-Cruz and J.J.~Toscano, 
             \Journal{\PRD}{62}{116005}{2002};
        J.L.~Diaz-Cruz, R.~Noriega-Papaqui, and A.~Rosado, 
             \Journal{\PRD}{71}{015014}{2005}. 

\bibitem{Iltan}
        E.O.~Iltan, 
       \Journal{JHEP}{0402}{065}{2004};
       \Journal{JHEP}{0408}{020}{2004}; 
       hep-ph/0504013.

\bibitem{BHHS}
        D.~Black, T.~Han, H.-J.~He, and M.~Sher,
        \Journal{\PRD}{66}{053002}{2002}.

\bibitem{bsmutau}
        D.~Guetta, J.M.~Mira, and E.~Nardi, 
        \Journal{\PRD}{59}{034019}{1999}.

\bibitem{super-B}
        A.G.~Akeroyd {\em et al.}, 
        {\it Super KEKB Letter of Intent}, 
        KEK Report 04-4, arXiv:hep-ex/0406071.

\bibitem{Pilaftsis} 
        A.~Pilaftsis, 
        \Journal{\PLB}{285}{68}{1992}.


\bibitem{Herrero}
        E.~Arganda, A.M.~Curiel, M.J.~Herrero, and D.~Temes,
        \Journal{\PRD}{71}{035011}{2005}.

\bibitem{kot}  S.~Kanemura, T.~Ota, K.~Tsumura, hep-ph/0505191.

\bibitem{Assamagan}
        K.A.~Assamagan, A.~Deandrea, and P.-A.~Delsart,
        \Journal{\PRD}{67}{035001}{2003}.

\bibitem{Osaka}
        S.~Kanemura, K.~Matsuda, T.~Ota, T.~Shindou, E.~Takasugi, and
        K.~Tsumura,     \Journal{\PLB}{599}{83}{2004}.

\bibitem{muon-collider}
        M.~Sher, \Journal{\PLB}{487}{151}{2000};
        U.~Cotti, M.~Pineda, and G.~Tavares-Velasco,
        arXiv:hep-ph/0501162. 

\bibitem{Sher-turan}
        M.~Sher and I.~Turan, \Journal{\PRD}{69}{017302}{2004}.

\bibitem{mutauDIS}
       S.~Kanemura, Y.~Kuno, M.~Kuze, and T.~Ota, 
       \Journal{\PLB}{607}{165}{2005}; 
       S.~Kanemura, Y.~Kuno, M.~Kuze, T.~Ota, and T.~Takai, 
       in preparation.      

\bibitem{VS}
        N.G.~Deshpande and E.~Ma,
        \Journal{\PRD}{18}{2574}{1978};
        S.~Kanemura, T.~Kasai, and Y.~Okada,
        \Journal{\PLB}{471}{182}{1999};
        S.~Nie and M.~Sher,
        \Journal{\PLB}{449}{89}{1999}.

\bibitem{PU-1}
        B.W.~Lee, C.~Quigg, and H.B.~Thacker,
        \Journal{\PRL}{38}{883}{1977};
        \Journal{\PRD}{16}{1519}{1977}.

\bibitem{PU-2}
        S.~Kanemura, T.~Kubota, and E.~Takasugi,
        \Journal{\PLB}{313}{155}{1993}.

\bibitem{LEP}
LEP Electroweak Working Group, 
                  {\it http://lepewwg.web.cern.ch/LEPEWWG/}. 

\bibitem{smlike1}
        I.F.~Ginzburg, M.~Krawczyk,  and P.~Osland, 
        arXiv:hep-ph/0101208;    
        {\em Nucl.Instrum.Meth.} {\bf A472} {149} {(2001)}.

\bibitem{bsg-ex}
        P.~Koppenburg {\it et al.}, the Belle Collaboration, 
        \Journal{\PRL}{93}{061803}{2004}.

\bibitem{smlike2}
        S.~Kanemura, S.~Kiyoura, Y.~Okada, E.~Senaha, and C.-P.~Yuan, 
        \Journal{\PLB}{558}{157}{2003}; 
        S.~Kanemura, Y.~Okada, E.~Senaha, and C.-P.~Yuan,
        \Journal{\PRD}{70}{115002}{2004}.

\bibitem{HHG} J.F.~Gunion, H.E.~Haber, G.~Kane, and S.~Dawson, 
        {\it The Higgs Hunter's Guide}, 
         Perseus Publishing, Cambridge, MA, 1990.

\bibitem{MSSMRN}
 F.~Borzumati and A.~Masiero, 
        \Journal{\PRL}{57}{961}{1986};
 J.~Hisano, T.~Moroi, K.~Tobe, M.~Yamaguchi, and T.~Yanagida,
        \Journal{\PLB}{357}{579}{1995};
 J.~Hisano, T.~Moroi, K.~Tobe, and M.~Yamaguchi,
        \Journal{\PRD}{53}{2442}{1996}.

\bibitem{Zee}
        A.~Zee, 
        \Journal{\PLB}{93}{389}{1980};
        Erratum \Journal{ibid}{95}{461}{1980}; 
        S.~Kanemura, T.~Kasai, G.-L.~Lin, Y.~Okada,  
        J.-J. Tseng, and C.-P. Yuan, 
        \Journal{\PRD}{64}{053007}{2001};
        K.~Cheung and O.~Seto, \Journal{\PRD}{69}{113009}{2004}. 

\bibitem{GH}
        J.F.~Gunion and H.E.~Haber, \Journal{\PRD}{67}{075019}{2003}. 

\bibitem{paradisi} 
        P.~Paradisi, hep-ph/0508054.




\end{thebibliography}
\end{document}